\documentclass[conference]{IEEEtran}
\IEEEoverridecommandlockouts
\def\BibTeX{{\rm B\kern-.05em{\sc i\kern-.025em b}\kern-.08em
    T\kern-.1667em\lower.7ex\hbox{E}\kern-.125emX}}

\IEEEoverridecommandlockouts

\def\BibTeX{{\rm B\kern-.05em{\sc i\kern-.025em b}\kern-.08em
    T\kern-.1667em\lower.7ex\hbox{E}\kern-.125emX}}

\usepackage{mathtools}
\usepackage[nodisplayskipstretch]{setspace}
\usepackage[utf8x]{inputenc}

\newcommand\bl[1]{\mathbf{#1}}
\def\blue{\textcolor{blue}}
\def\red{\textcolor{red}}

\usepackage{subfigure}
\usepackage{moreverb}
\usepackage{epsfig}
\usepackage{amsmath,amssymb,bm,amsthm,mathrsfs,dsfont}
\usepackage{fancyhdr}
\usepackage{adjustbox,lipsum}
\usepackage[font={small}]{caption}
\usepackage{color,soul}
\usepackage{graphicx}
\usepackage{amsfonts}
\usepackage{cite}
\usepackage{microtype}

\usepackage{nomencl}
\makenomenclature

\usepackage{algorithmicx}
\usepackage{algpseudocode}
\usepackage[norelsize, linesnumbered, ruled, vlined, lined, boxed, commentsnumbered]{algorithm2e}

\begin{document}
\title{ 
Constrained Multimodal Sensing-Aided Communications: A Dynamic Beamforming Design
}
\author{\IEEEauthorblockN{
 Abolfazl Zakeri\IEEEauthorrefmark{1}, Nhan~Thanh~Nguyen\IEEEauthorrefmark{1},
Ahmed Alkhateeb\IEEEauthorrefmark{2},
 and Markku Juntti\IEEEauthorrefmark{1}}
 \vspace{-1 em }
 \\
 \IEEEauthorrefmark{1}\normalsize 
CWC-RT, University of Oulu, Finland,
 Email: 
\{abolfazl.zakeri,\,nhan.nguyen,\,markku.juntti\}@oulu.fi
\\
\IEEEauthorrefmark{2}The School of Electrical, Computer, and Energy Engineering, Arizona State University, USA, Email: alkhateeb@asu.edu
}

\maketitle
	\begin{abstract}
Using multimodal sensory data can enhance communications systems by reducing the overhead and latency in beam training. 
However, processing such data incurs high computational complexity, and continuous sensing results in significant power and bandwidth consumption. This gives rise to a tradeoff between the (multimodal) sensing data acquisition rate and communications performance. In this work, we develop a constrained multimodal sensing-aided communications framework where dynamic sensing and beamforming are performed under a sensing budget. Specifically, we formulate an optimization problem that maximizes the average received signal-to-noise ratio (SNR) of user equipment, subject to constraints on the average number of sensing actions and power budget. Using the Saleh-Valenzuela mmWave channel model, we construct the channel primarily based on position information obtained via multimodal sensing. Stricter sensing constraints reduce the availability of position data, leading to degraded channel estimation and thus lower performance. We apply Lyapunov optimization to solve the problem and derive a dynamic sensing and beamforming algorithm. Numerical evaluations on the DeepSense and Raymobtime datasets show that \textit{halving} sensing times leads to only up to 7.7\% loss in average~SNR.
\end{abstract}
\section{Introduction} 
Leveraging multi-modal sensory data collected from visual, LiDAR, and radar sensors for communications, referred to as ``multimodal sensing-aided communications", has recently gained increasing attention due to the enhanced environmental perception and situational awareness, and thereby enabling more informed and adaptive decision-making \cite{Ahmed_deepsense, R_Heath_mag, multimodality_mmwave_mag}.
Its applications span various domains, including vehicular communications, positioning, and healthcare.
Among the primary communications tasks where multimodal sensing has seen growing use is beamforming design \cite{Ahmed_seman_dis, RHeath_tran_multiuser-beamforming, multimodality_connected_vechile}.
It offers significant potential in reducing beam training overhead in high-frequency systems such as mmWave
 \cite{RHeath_tran_multiuser-beamforming, Ahmed_seman_dis}, and in improving beam alignment in connected vehicle scenarios \cite{multimodality_connected_vechile}.
This advantage becomes particularly prominent in scenarios with highly mobile users, where proactive line-of-sight (LoS) link prediction and future beam selection are essential.
\\\indent
Building on this premise, several studies have explored multimodal sensing in communications systems, spanning applications such as beam selection in different vehicular networks \cite{Ahmed_vision, multimodality_connected_vechile,latin_amr_RH,transformer_sensing} to digital twin frameworks \cite{salehihikouei2024leveraging}. The majority of existing research has focused on various beamforming problems, particularly codebook-based approaches, in mmWave multiple-input multiple-output (MIMO) systems. Collectively, these works demonstrate that multimodal sensing can significantly reduce beam training overhead while maintaining, or even improving, beamforming performance.
\\\indent
Nonetheless, most existing studies overlook practical limitations related to the continuous collection and processing of sensed information, such as increased computational complexity and limited scalability. Moreover, continuous multimodal data acquisition, hereafter referred to as ``sensing'',\footnote{In this paper, we use the term ``sensing'' to denote the process of collecting and processing multimodal sensory data related to the communications environment. This usage differs from conventional radar sensing applications such as target detection and localization.} incurs excessive power and bandwidth consumption. This introduces a tradeoff between the sensing rate and communications performance, which remains unexplored in the literature. To address this gap, we propose a constrained sensing-aided communications framework aimed at scenarios where sensing can only be performed a limited number of times, reflecting the resource and complexity constraints associated with the sensing process. Our approach offers an efficient alternative to continuous, resource-intensive data acquisition across all deployed (multimodal) sensors, which may not be possible in practice.
\\\indent
We consider a setup with a fixed base station (BS) equipped with multiple antennas as a transmitter and a single-antenna mobile user equipment (UE). We construct the channel using the Saleh-Valenzuela channel model, where a dominant LoS path, determined by the UE's position, is the primary component.
 However, maintaining continuous availability of position information at the BS incurs resource consumption and may not always be feasible in practice. 
To address this, we formulate an optimization problem that maximizes the average received signal-to-noise ratio (SNR) of the UE, subject to a limit on the average number of time slots during which position information is available at the BS, as well as a transmit power budget. A key tradeoff here is that fewer position information updates, i.e., stricter the sensing constraint, reduce the availability of position data at the BS, leading to degraded channel estimation and consequently performance loss. We employ the Lyapunov optimization to solve the average problem and propose a dynamic sensing and beamforming algorithm. 
We conduct the simulations based on DeepSense \cite{Ahmed_deepsense} and Raymobtime \cite{RHeath_Raymobtime} datasets, observing only up to 7.7\% loss in average received SNR by UE, while using position information only \textit{half} the time over the dataset duration.
\section{System Model and Problem Formulation}
\subsection{System Model}
We consider a downlink communications system with a BS, located at a fixed position, and a mobile UE.
The BS is equipped with a uniform linear array (ULA) with $N$ antennas, and the UE is equipped with a single-antenna receiver. At time slot $t=1,2,\dots$, the BS transmits the data signals to the UE using the beamforming vector $\mathbf{w}(t)\in\mathbb{C}^{N\times 1}$.
\\\indent
Let  $\mathbf{h}(t)\in\mathbb{C}^{N\times 1}$ denote the (downlink) channel from the BS to the UE at time slot $t$. Then the received signal at slot $t$ at the UE is given by
\begin{align}
    y(t) = \mathbf{h}^{\textsf{H}}(t)\bold{w}(t)+n(t),
\end{align}
where ${s}(t)$ is the transmit (data) signal to the UE, ${\Bbb{E}\{|{{s}(t)}|^2\}=1}$,
and $n(t)\in \Bbb{C}$ is additive white Gaussian noise (AWGN) drawn from the distribution $\mathcal{CN}(0,\sigma^2)$, with $\sigma^2$ being the noise power at the UE's receiver. Furthermore, the received SNR at the UE at time slot $t$ is given by 
\begin{equation}
\dfrac{|\mathbf{h}^{\textsf{H}}(t)\mathbf{w}(t)|^2}{\sigma^2}\cdot
\end{equation}
\indent
Our goal is to design the beamfoming vector $\mathbf{w}(t)$ for each time $t$ that maximizes the average received SNR of the UE.
Given the channel $\mathbf{h}(t)$, this is a relatively simple task. More precisely, an optimal beamformer is a matched filter, i.e., aligning the beamforming vector with the channel, which is commonly known as a maximum ratio transmission (MRT) beamformer. However, the main challenge here is that obtaining the channel information at all times requires excessive signaling overhead, additional latency, and resource consumption. 
To overcome this, similarly to, e.g., \cite{R_Heath_mag,latin_amr_RH,Ahmed_vision}, we propose the idea of using multimodal sensing for dynamic beamforming.\footnote{We use the term ``multimodal sensing" to refer to the general functionality of environmental awareness. However, our current design leverages only monomodal sensory data, specifically, position information, for dynamic beamforming. The extension to integrate multiple data modalities is left for future work.}
\\\indent 
Our multimodal sensing approach in this paper is to construct the channel based on the position information first and then derive the beamforming for the obtained channel. 
The primary reason for considering the position modality is twofold: (1) it is relatively low-cost compared to other modalities such as LiDAR or visual camera images, and (2) it provides sufficient information to effectively evaluate our proposed constrained multimodal sensing framework.
Next, we introduce our channel model first, and then the problem formulation.
\\\indent
\textit{Channel Modeling:}
We consider a widely used Saleh-Valenzuela mm-Wave channel model to construct the channel primarily based on the dominant LoS path determined by the position information. 
The channel vector $\mathbf{h}(t)$ is given by 
\begin{align}\label{eq_chnl_}
    \mathbf{h}(t) = \mathbf{h}_{\textsf{LoS}}(t) + \mathbf{h}_{\mathsf{NLoS}}(t) = \sum_{l=0}^{L} \beta_l(t) \mathbf{a}(\theta_l(t)) ,
\end{align}
where $\mathbf{h}_{\textsf{LoS}}(t)$ and $\mathbf{h}_{\mathsf{NLoS}}(t)$ are respectively the LoS and non-LoS (NLoS) components of the channel, and $L$ is the number of multipath components. Moreover, $\beta_l(t)$ represents the path gain, $\theta_l(t)$ is the angle of departure (AoD), and $\mathbf{a}(\theta_l(t))$ is the steering vector of the $l$-th path, with $l=0$ corresponding to the LoS component. 
Given the ULA antenna architecture deployed in the BS and assuming half wavelength antenna spacing, the steering vector is given by
\begin{align}
   \mathbf{a}(\theta_l(t)) = \frac{1}{\sqrt{N}} \left[1, e^{j \pi \sin(\theta_l(t))}, \dots, e^{j (N-1)\pi \sin(\theta_l(t))}\right]^{\mathrm{T}}.
 \end{align}
We assume that $\beta_l(t)$ in \eqref{eq_chnl_} is always normalized with the path gain of LoS, i.e., $\beta_0(t) =1$, and the values of $\beta_l(t) \ll \ 1$, $l=1,\dots, L-1$, are randomly generated 
relative to the LoS' path gain. 
The position information determines the AoD of the LoS path, $\theta_0(t)$, and the AoD of the NLoS paths are generated randomly.     
\subsection{Problem Formulation} 
Given the channel model derived above, which relies primarily on position information, the next step is to derive the optimal beamforming vector. However, continuously acquiring position information at the BS in every time slot is costly or may not always be feasible in practice due to factors such as potential sensor failures. This limits the channel acquisition and thus causes performance loss. Consequently, a tradeoff exists between the frequency of position updates and the resulting beamforming performance. Below, we formulate this tradeoff problem.
\\\indent 
Let \( x(t) \in \{0,1\} \) be a binary (multimodal) sensing decision variable indicating whether the UE’s position information at time slot \( t \) is available at the BS; specifically, \( x(t) = 1 \) if the BS obtains the UE’s position at slot \( t \), and \( x(t) = 0 \) otherwise. Accordingly, the decision on $x(t)$ determines the availability of the UE's position information at the BS, and consequently, the availability of the corresponding channel at the BS, i.e., $\mathbf{h}_{\mathsf{LoS}}(t)$. 
Notice that because the position information is the only available information at the BS, the (full) channel 
$\mathbf{h}(t)$ is not accessible by the BS for beamforming.
\\\indent
 When \( x(t) = 1 \), the LoS channel component based on the \textit{current} position information is available at the BS. Conversely, when \( x(t) = 0 \), we assume that the LoS channel component, based on the \textit{most recently} available position information, is given at the BS. Let \( \tilde{\mathbf{h}}(t) \) denote the \textit{available} channel at the BS at each time slot~\( t \),\footnote{This can also be interpreted as the \textit{estimated} channel at the BS.} which is given by
\begin{align}\label{eq_hhat_x0}
   \tilde{\mathbf{h}}(t) = \begin{cases}
       \mathbf{h}_{\mathsf{LoS}}(t), & ~\text{if}~x(t)=1,
       \\
       \mathbf{h}_{\mathsf{old}}(t), & ~\text{if}~x(t)=0,
   \end{cases}
\end{align}
where  $\mathbf{h}_{\mathsf{old}}(t) = \mathbf{h}_{\mathsf{LoS}}(t')$ for the latest time $t'<t$ that ${x(t')=1}$.
\\\indent
Based on these definitions, we aim to optimize the time-specific beamforming vectors and sensing decisions, i.e., $\{\mathbf{w}(t), x(t)\}_{t=1,2,\ldots}$, to maximize the average received SNR at the UE, subject to constraints on the average number of sensing times and the transmit power budget at the BS.
This problem is formulated as:
\begin{subequations}
       \label{op_1}
       \begin{align}
     \underset{\{\mathbf{w}(t), x(t)\}_{t=1,2,\ldots}}{\mbox{maximize}}~~~   & \label{eq_obj_fun}
           \limsup_{T\rightarrow \infty } \dfrac{1}{T} \sum_{t=1}^{T} \Bbb{E} \{|\mathbf{h}^{\textsf{H}}(t)\mathbf{w}(t)|^2\}
           \\
        		\mbox{subject to}~~~~~~~~ &  
                \label{eq_cons_sensing}
               \limsup_{T\rightarrow \infty } \dfrac{1}{T} \sum_{t=1}^{T} \Bbb{E} \{x(t)\} \leq \alpha,
                \\ &
x(t)\in\{0,\,1\},  ~\forall\, t,
                \\&
                \label{eq_cons_power}
                 \|\mathbf{w}(t)\|^2 \leq P_{\mathrm{max}}, ~\forall\, t.
                \end{align}
        		\end{subequations} 
              In \eqref{eq_obj_fun}, operation $\mathbb{E}\{\cdot\}$ denotes the expectation taken over the potential randomness in the channel and the sensing and beamforming decisions made based on the available channel at the BS $\tilde{\mathbf{h}}(t)$.
Moreover, $\alpha\in(0,1]$ represents the limit on the average number of times (the real-time) position information is available. The more frequently 
$x(t)=1$, the more accurate the available channel, hence the better beamforming performance. However, constraint~\eqref{eq_cons_sensing} essentially limits the number of times multimodal sensing is performed, and consequently, the availability of position information.
                Thus, problem~\eqref{op_1} broadly studies a trade-off between the \textit{multimodal sensing cost} in terms of complexity and/or resource usage and the \textit{communications performance}. 

 \section{Proposed Solution to Problem \eqref{op_1}}
To solve problem~\eqref{op_1}, we employ Lyapunov optimization, specifically the drift-plus-penalty method~\cite{Neely_Sch}, and develop an algorithm to jointly optimize 
the transmit beamforming and the time slots to perform (multimodal) sensing. 
This algorithm provides a low-complexity heuristic solution to the original problem and does not require prior knowledge of system dynamics, such as the UE's mobility pattern. 
The main idea of the drift-plus-penalty method is to enforce the average constraint~\eqref{eq_cons_sensing} through queue stability and to transform the original average problem~\eqref{op_1} into a sequence of per-slot optimization problems. 
\\\indent
Let $Q(t)$ denote the virtual queue associated with constraint~\eqref{eq_cons_sensing} in slot $t$ which evolves as 
\begin{equation}\label{eq_virtualQueue}
    Q(t+1) = \max [Q(t) + x(t) -\alpha,0]. 
\end{equation}
The process $Q(t)$ can be seen as a queue with service rate $x(t)$ and arrival rate $\alpha$.
By \cite[Ch. 2]{Neely_Sch}, the time average constraint~\eqref{eq_cons_sensing} is satisfied when the queue is strongly stable, i.e., $\limsup_{T\rightarrow \infty} \frac{1}{T}\sum_t^T \Bbb{E}\{Q(t)\} < \infty$. Next, we define
the Lyapunov function and its drift to account for the queue stability and proceed with the drift-plus-penalty method. 

Let $L(Q(t))=\frac{1}{2}Q^2(t)$ be the quadratic Lyapunov function \cite[Ch. 3]{Neely_Sch}. By minimizing the expected change of the Lyapunov function
from one slot to the next, the virtual queue can be stabilized
\cite[Ch. 3]{Neely_Sch}. Let $S(t)\triangleq \{Q(t),\tilde{\mathbf{h}}(t)\}$ denote the network state
in slot~$t$. The one-slot conditional Lyapunov drift, denoted
by $\Delta(t)$, is the expected change in the Lyapunov
function over one slot given the current system state $S(t)$.
Accordingly, $\Delta(t)$ is defined as \cite[Eq. 3.13]{Neely_Sch}
\begin{equation}\label{eq_drift}
    \Delta(t) = \Bbb{E} \{L(Q(t+1)) - L(Q(t))\,|\,S(t)\}.
\end{equation}
	Applying the drift-plus-penalty method, we need to find $x(t)$ and  $\mathbf{w}(t)$ every time slot $t$ that  minimizes a bound on the following drift-plus-penalty function 
    \begin{equation}\label{eq_dpp_func}
    \Delta(t) -  V \Bbb{E}\{|\mathbf{h}^{\mathsf{H}}(t)\mathbf{w}(t)\,|\,S(t)\}  ,
    \end{equation}
    subject to the power constraint \eqref{eq_cons_power}, where $V$ is a non-negative parameter
    chosen to desirably adjust a trade-off between the size of the virtual queue and the objective function of \eqref{op_1}.
\\\indent
Optimizing directly \eqref{eq_dpp_func}  is difficult owing to function $\max[\cdot]$ in the virtual queue evolution in \eqref{eq_virtualQueue}.
Leveraging the fact that for any $c \ge 0, b \ge 0, A \ge 0$, we have \cite[p. 33]{Neely_Sch}
$$(\max[c-b,0]+A)^2 \le c^2+A^2+b^2+2c(A-b),$$ we can derive the following upper-bound for $\Delta(t)$:
\setlength{\abovedisplayskip}{5pt}
\setlength{\belowdisplayskip}{5pt}
\begin{equation}
    \Delta(t) \le C + Q(t)\alpha + \Bbb{E}\{Q(t)x(t)\,|\,S(t)\},
\end{equation}
where $C$ is a positive constant. 
\\\indent 
Following the standard procedure of the drift-plus-penalty method, we use the approach of opportunistically minimizing an expectation to optimize the upper-bound of the drift-plus-penalty function \cite[Ch. 3]{Neely_Sch}. Noting that the constant terms in the drift-plus-penalty do not impact the solution, to obtain our dynamic sensing and beamforming algorithm, we now aim to solve the following \textit{per-slot} optimization problem:
\begin{subequations}
       \label{op_perslot}
       \begin{align}
        \underset{\mathbf{w}(t),\,x(t)}{\mbox{maximize}}~~~   &\label{eq_obj_prslot}
       V |\mathbf{h}^{\mathsf{H}}(t)\mathbf{w}(t)| - Q(t)x(t)
           \\
        		\mbox{subject to}~~~ 
                & x(t)\in\{0,\,1\},
               \\ & \label{eq_cons_power_perslot}
                             \|\bl{w}(t)\|^2 \leq P_{\mathrm{max}}, 
                \end{align}
        		\end{subequations}
                 where $Q(t)$ is the virtual queue at the time slot $t$. Notice that in the above problem~\eqref{op_perslot} the channel $\mathbf{h}(t)$ is not available and $x(t)$ determines the available channel based on which the beamforming vector is designed. 
\\\indent
To solve the above problem, we note that given $x(t)$, an optimal beamformer is the matched filter given by
\begin{align}
    \label{eq_bmf_opt}
    \mathbf{w}^{\star}(t) = \sqrt{P_{\max}} \dfrac{\tilde{\mathbf{h}}(t)}{\|\tilde{\mathbf{h}}(t)\|}\cdot
\end{align}
Because $x(t)$ is a binary decision, we apply an exhaustive search to determine it and then obtain an optimal beamforming according to~\eqref{eq_bmf_opt}. Details of the proposed dynamic sensing and beamforming based on the drift-plus-penalty method are given in Alg.~\ref{alg_sen_bmf}.
\begin{algorithm}[t!]
   \caption{Dynamic Algorithm to Solve Problem~\eqref{op_1}}
   \label{alg_sen_bmf}
   \SetKwInOut{Inputi}{Initialize}
   \SetKwInOut{Output}{Output}
   \SetKwComment{Comment}{/* }{ */}
   \setlength{\AlCapSkip}{1em}

   \Inputi{
       Set $t=0$, control parameter $V$; initialize $Q(0)=0$. 
   }

   \For{each time slot $t$}{
        Set $x(t)=1$ and obtain channel estimate $\tilde{\mathbf{h}}(t)$ using~\eqref{eq_hhat_x0}\;
        Compute $\mathbf{w}_1(t)$ by~\eqref{eq_bmf_opt} given $\tilde{\mathbf{h}}(t)$\;
        Compute objective $\mathsf{obj}_1 = V |\mathbf{h}^{\mathsf{H}}(t)\mathbf{w}_1(t)|^2 - Q(t)x(t)$\;

        Set $x(t)=0$ and obtain channel estimate $\tilde{\mathbf{h}}(t)$ using~\eqref{eq_hhat_x0}\;
        Compute $\mathbf{w}_0(t)$ by~\eqref{eq_bmf_opt} given $\tilde{\mathbf{h}}(t)$\;
        Compute objective $\mathsf{obj}_0 = V |\mathbf{h}^{\mathsf{H}}(t)\mathbf{w}_0(t)|^2 - Q(t)x(t)$\;

        \eIf{$\mathsf{obj}_1 \ge \mathsf{obj}_0$}{
            Set $x^{\star}(t) = 1$, $\mathbf{w}^{\star}(t) = \mathbf{w}_1(t)$\;
        }{
            Set $x^{\star}(t) = 0$, $\mathbf{w}^{\star}(t) = \mathbf{w}_0(t)$\;
        }

        Update $Q(t+1)$ using~\eqref{eq_virtualQueue}, and update $\mathbf{h}_{\mathsf{old}}(t+1)$\;
   }
\end{algorithm}
\\\indent
Finally, because the UE's received SNR remains finite for any beamforming design owing to the power budget, one can show that, for a finite value of $V$, the derived Alg.~\ref{alg_sen_bmf} is guaranteed to return a strongly stable virtual queue; this, in turn, implies the satisfaction of the average sensing constraint~\eqref{eq_cons_sensing}. This is further verified via simulation results shown in the next section~(Fig.~\ref{fig_avrsensing_vs_alpha}). 

\vspace{-1.5  em}
\section{Numerical Results}\label{sec_numres} 
\begin{figure}[t!]
\centering
\subfigure[DeepSense dataset, with $V=1$] 
{
\includegraphics[width=0.3155\textwidth]{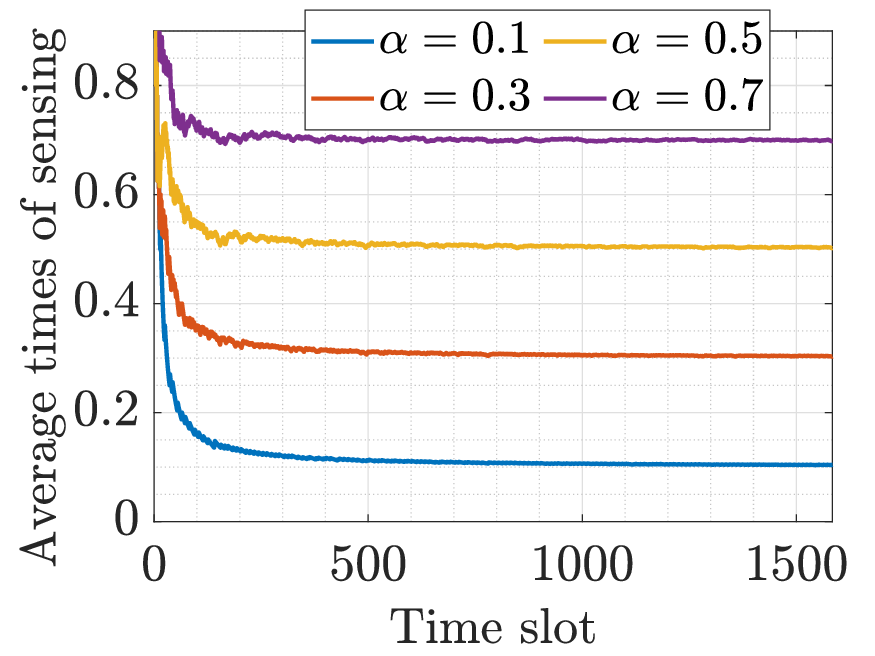}
\label{fig_avrsentime_alpha_DeepSen}
}
\subfigure[Raymobtime dataset, with $V=10$ ]{
\includegraphics[width=0.32\textwidth]{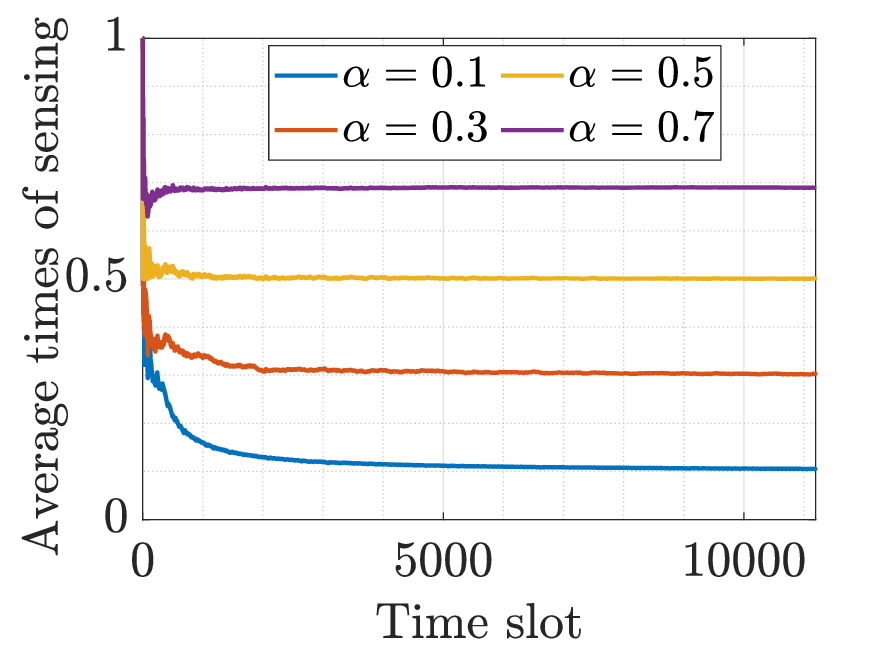 }
\label{fig_avrsentime_alpha_Ray}
}
\caption{Satisfaction of constraint~\eqref{eq_cons_sensing} by the derived Alg.~\ref{alg_sen_bmf} for different datasets, $\alpha$ is the sensing budget.
}
\label{fig_avrsensing_vs_alpha}
\vspace{-2 em}
\end{figure}

\begin{figure}[t!]
\centering
\subfigure[Average~SNR vs. $\alpha$] 
{
\includegraphics[width=0.33\textwidth]{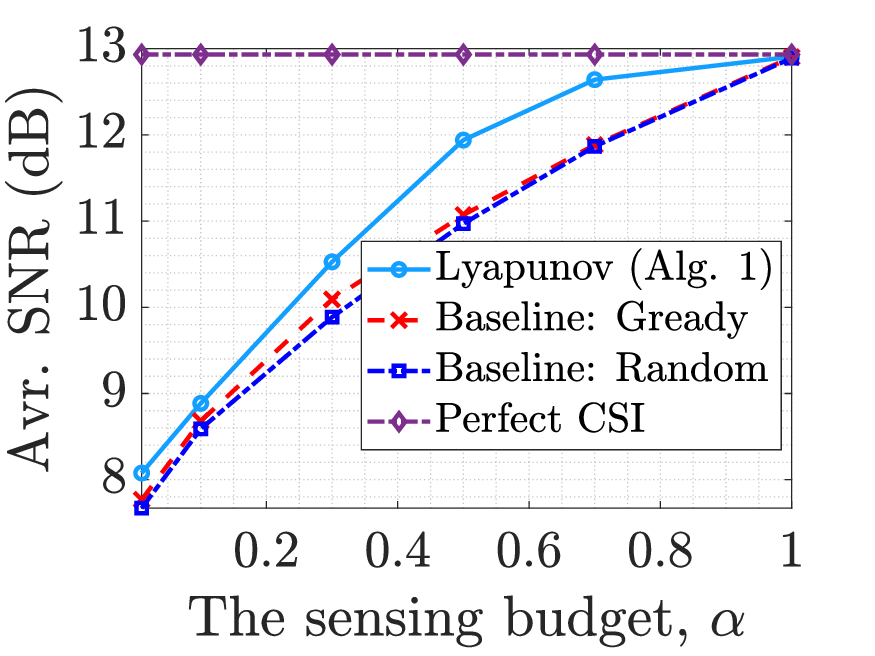}
\label{fig_avrSNR_alpha_DeepSen}
}
\subfigure[Average~SNR vs. $P_{\max}$]{
\includegraphics[width=0.33\textwidth]{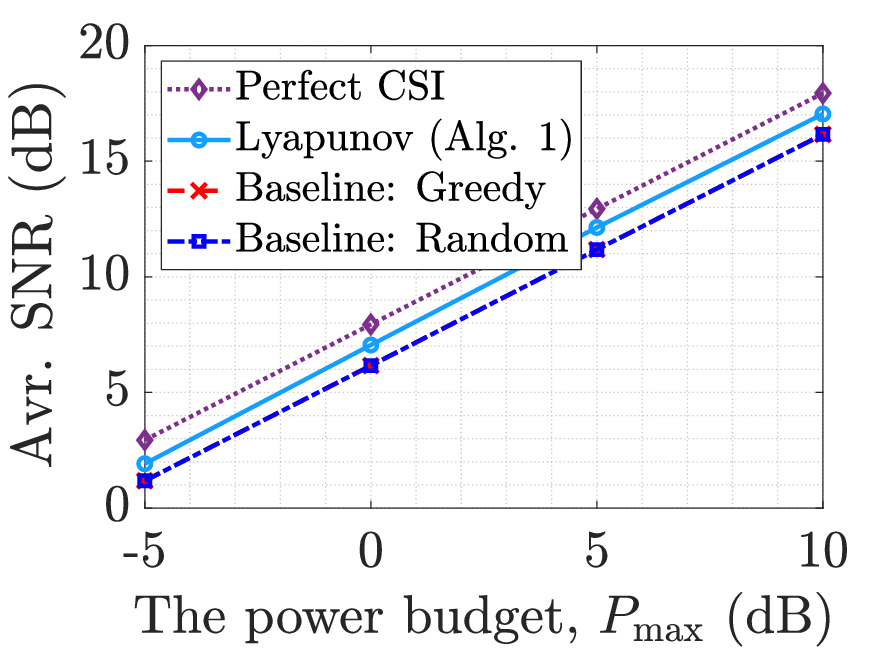 }
\label{fig_avrSNR_P_max_DeepSen}
}
\caption{ Average SNR comparison between different algorithms for DeepSense dataset, where $V=1$.
}
\label{fig_avrSNR_DeepSen}
\vspace{-2 mm}
\end{figure}

This section provides simulation results to demonstrate the effectiveness of the derived algorithm and the impact of some critical parameters on performance. 
For the benchmarking we consider the following algorithms: (i) \textit{greedy-based sensing} according to which, at each time $t$,  the variable $x(t)=1$ if $\bar{\alpha}(t) \le \alpha$, where $\bar{\alpha}(t)$ is the averaged value of $\{x(0),\dots,x(t)\}$; and (ii) \textit{randomized sensing} by which we determine $x(t)$ randomly provided that constraint~\eqref{eq_cons_sensing} is satisfied. We further consider the case where the constructed channel $\mathbf{h}(t)$ is available at the BS in all time slots, termed ``\textit{perfect} channel state information (\textit{CSI})", which gives an upper-bound; notice that this is not necessarily equal to the true channel in the real-world associated with the used datasets scenarios in this paper.
\\\indent
For the position information utilized in the channel construction, we consider:
(1) DeepSense scenario 5 dataset \cite{Ahmed_deepsense}, and (2) Raymbtime $\mathsf{s008}$ dataset available in \cite{RHeath_Raymobtime}, used in \cite{ssp_sensing}.
Unless otherwise stated, we set $N=6$, $L = 6$, $\sigma^2=1$, and $P_{\max}=5$ dB. 
The phases of five multipaths are generated uniformly at random, and the corresponding normalized path gains are uniformly drawn from the set $\{10^{-1},\dots, 10^{-5}\}$.
Furthermore, it is noteworthy that the period over which we run the algorithms and take the average is limited to the number of epochs in each dataset. 
\\\indent
 First, in Fig.~\ref{fig_avrsensing_vs_alpha}, we plot the average number of slots in which real-time position data is available at the BS, i.e., \(x(t) = 1\), referred to as the average number of sensing slots, for various values of the sensing budget \(\alpha\) in \eqref{eq_cons_sensing}. This figure is aimed to verify the satisfaction of the average constraint~\eqref{eq_cons_sensing} in the main problem by Alg.~\ref{alg_sen_bmf}. The figure demonstrates that the derived Alg.~\ref{alg_sen_bmf} indeed satisfies the average sensing constraint for all values of \(\alpha\) and across both the datasets.
\vspace{0.1em}
\\\indent
Fig.~\ref{fig_avrSNR_DeepSen} depicts the average received SNR as a function of the sensing and power budgets for different algorithms for the DeepSense dataset. The first observation from Fig.~\ref{fig_avrSNR_alpha_DeepSen} is that increasing the sensing budget, which means more real-time access to the position information, improves the average SNR. It is important to note that the impact of $\alpha$, as a sensing frequency here, is dependent on the time-scale of the collected data; the higher the sensing frequency is for the data collection in the real world, the less sensitive the results should be to the variations of $\alpha$. 
Additionally, Fig.~\ref{fig_avrSNR_P_max_DeepSen} shows how the BS power budget impacts average SNR for different algorithms for fixed $\alpha=0.5$. We observe that the proposed Alg.~\ref{alg_sen_bmf} outperforms the other benchmarks while retaining almost comparable performance compared to the ``perfect CSI'' case. The trend is as expected due to the fact that the more transmit power, the stronger the received signal, regardless of the beamforming direction. 

\begin{figure}[t!]
\centering
\subfigure[Average~SNR vs.~$\alpha$ for~$P_{\max}=5$~~]
{
\includegraphics[width=0.33\textwidth]{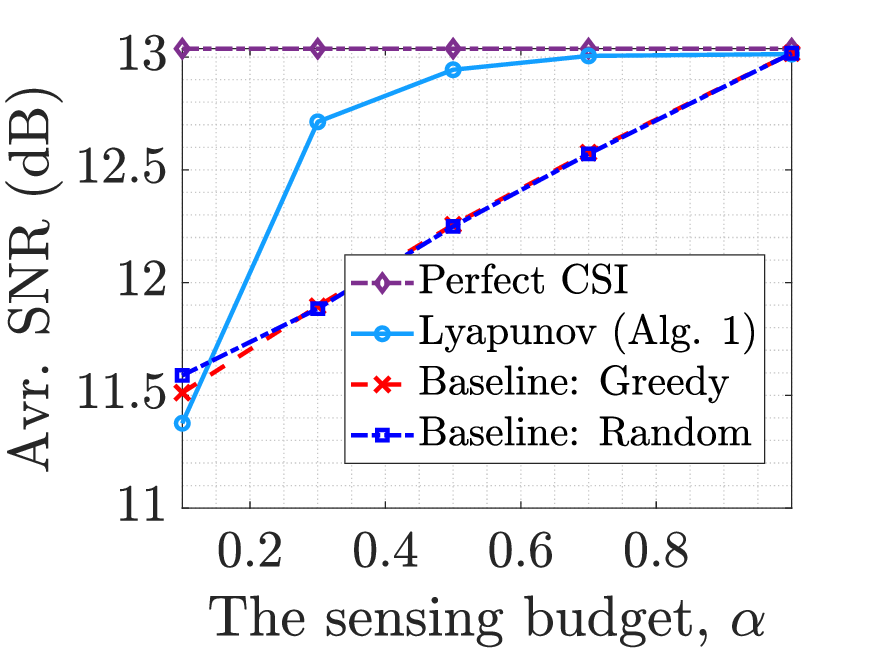}
\label{fig_avrSNR_alpha_Raymob}
}
\centering
\subfigure[Average SNR vs. $P_{\max}$ for $\alpha=0.3$]{
\includegraphics[width=0.33\textwidth]{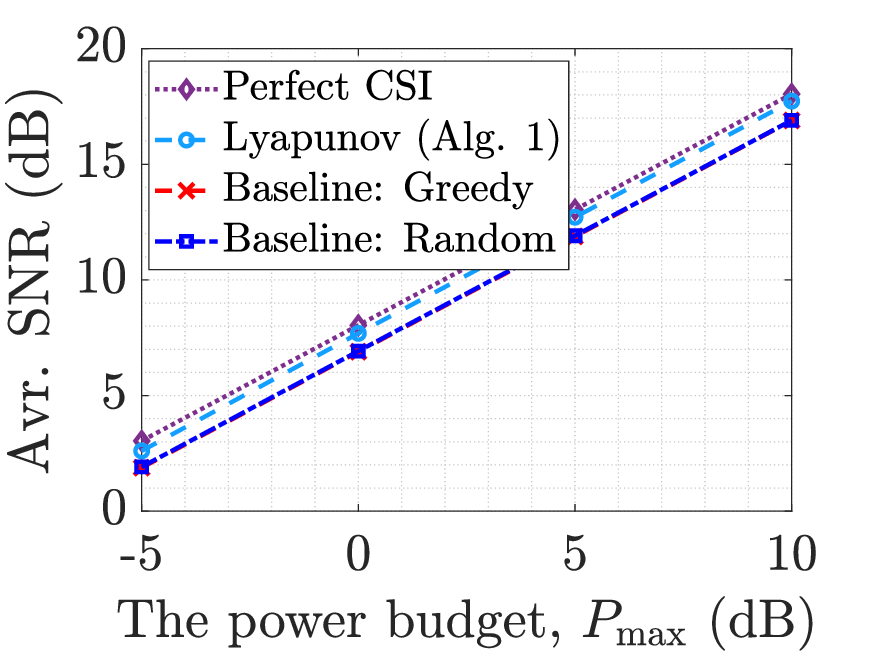 }
\label{fig_avrSNR_P_max_Raymob}
}
\caption{Average SNR comparison between different algorithms for Raymobtime dataset, where $V=10$. 
}
\label{fig_avrSNR_Raymob}
\vspace{-3 mm}
\end{figure}

The same analysis as in Fig.~\ref{fig_avrSNR_DeepSen} is conducted in Fig.~\ref{fig_avrSNR_Raymob}, this time using position information from the Raymobtime dataset. Overall, similar observations hold: both the sensing budget and power budget directly influence the performance of all algorithms. Notably, the proposed algorithms outperform the baseline methods, with a more pronounced performance gap compared to Fig.~\ref{fig_avrSNR_DeepSen}. Furthermore, for $\alpha \ge 0.5$, the performance of the proposed algorithm approaches that of perfect CSI, highlighting the effectiveness of our framework as a foundation for carefully designed sensing-aided communications under sensing/data collection constraints owing to resource limitations.

\vspace{-1 em}
\section{Conclusions }\label{sec_conl}
We addressed a tradeoff problem between (multimodal) sensing cost, represented by the number of time slots in which sensory data is available at the BS, and beamforming performance in a multimodal sensing-aided communications system. 
Stricter constraints on sensing reduce the number of position updates, which degrades channel estimation accuracy and, consequently, beamforming performance.
We formulated this as an optimization problem that aims to maximize the average SNR at the UE, subject to limits on the average number of times position data is available and the transmit power. To solve this problem, we applied Lyapunov optimization and derived a dynamic beamforming.

Simulation results using the DeepSense and Raymobtime datasets showed that reducing the use of position data for beamforming by 50\% leads to only a 7.7\%  drop in average SNR. This suggests that there is a potential for resource-efficient multimodal sensing algorithms that obtain desirable communications performance while reducing sensing overhead in terms of complexity and resource usage.

\bibliographystyle{ieeetr}
\bibliography{Bib_References/conf_short,
Bib_References/IEEEabrv,
Bib_References/Bibliography}

\end{document}